\documentclass{article}

\usepackage{PRIMEarxiv}

\usepackage[utf8]{inputenc} 
\usepackage[T1]{fontenc}    
\usepackage{hyperref}       
\usepackage{url}            
\usepackage{booktabs}       
\usepackage{amsfonts}       
\usepackage{nicefrac}       
\usepackage{microtype}      
\usepackage{amsmath}        
\usepackage{lipsum}
\usepackage{fancyhdr}       
\usepackage{graphicx}       
\graphicspath{{media/}}     

\usepackage{titlesec}
\usepackage{pgf-pie}

\titlespacing*{\section}
{0pt}        
{1.2ex}      
{0.8ex}      

\title{Cross-Vendor Sola
ISPM Benchmark:\\ Evaluating Agentic AI for Federated Identity Security Reasoning
}

\author{
\textbf{Eden Yavin},
  \textbf{Gal Engelberg},
  \textbf{Konstantin Koutsyi},
  \textbf{Leon Goldberg},
  \textbf{Gal Baron}\\
  Sola Security, Tel Aviv, Israel \\
  \texttt{Corresponding author: eden.y@sola.security}
}

\begin{document}
\maketitle

\begin{abstract}
The rapid proliferation of multi-cloud and SaaS platforms has transformed Identity Security Posture Management (ISPM) into a fundamentally cross-vendor challenge: critical misconfigurations and privilege escalation paths increasingly span multiple identity providers, infrastructure layers, and authentication systems that were never designed to interoperate. Existing evaluations focus on isolated single-platform environments and provide no means to assess whether an AI agent can reason across these fragmented boundaries. To address this gap, we introduce the \textit{Cross-Vendor Sola ISPM Benchmark}, a production-grade benchmark of 50 data-grounded tasks requiring multi-hop entity resolution and cross-system correlation across eight integrated enterprise platforms including AWS, Okta, Azure AD, and Google Workspace. Alongside the benchmark, we contribute an evaluation framework that measures not only final answer correctness but also evidentiary grounding, structural join fidelity, retrieval quality, and SQL equivalence. We evaluate the Sola AI Agent across five context configurations - from no injected metadata to full schema, graph, and retrieval context - using three frontier LLMs.
The results show that enriching the agent with structured relational context 
consistently improves answer correctness by approximately 34\% relatively, and reduces the average number of exploration queries by approximately 70\% across all 
tested models, with the largest gains driven by the inclusion of cross-vendor 
graph topology.
Our findings indicate that frontier LLMs possess substantial latent security reasoning capability, but their ability to perform reliable cross-vendor identity analysis is fundamentally constrained by the availability of explicit relational context for entity resolution and evidentiary grounding. Under full context, the best-performing configuration achieves an answer correctness score of $78\%$ while reducing complete failure to $4\%$.
\end{abstract}


\section{Introduction}\label{sec:intro}

The modern enterprise security perimeter has evolved from isolated platforms into a highly decentralized web of multi-cloud and Software-as-a-Service (SaaS) architectures, where identity serves as the primary control plane~\cite{RSA2023_ISPM}. In this environment, critical security risks increasingly reside at the intersection of diverse ecosystems spanning disparate Identity Providers (IdPs), Cloud Service Providers (CSPs), and productivity suites. Securing these federated boundaries requires reasoning across systems simultaneously - detecting risks such as MFA bypass vulnerabilities, dormant accounts propagating across tenants, or privilege escalation paths that originate in an IdP and terminate in a CSP. Despite the rapid integration of Large Language Models (LLMs) into security operations~\cite{GoogleCloud2025_ROI_AI_Security}, and a growing industry reliance on agentic AI to achieve the autonomous, cross-system reasoning these tasks demand, the field currently operates in a critical evaluation void: there is no standardized, data-grounded benchmark for assessing agentic AI on complex, multi-system Identity Security Posture Management (ISPM) workflows.

While significant progress has been made in benchmarking AI for discrete cybersecurity domains - such as Security Operations Center (SOC) investigations~\cite{ExCyTInBench_SecRL}, cyber threat intelligence synthesis~\cite{CTIBench, SEvenLLM_Bench}, and Role-Based Access Control (RBAC) policy comprehension~\cite{OrgAccess_Benchmark_RBAC_Reasoning} - the cross-vendor dimensions of ISPM remain entirely unmeasured. Organizations increasingly rely on automated systems to govern complex integrations, yet they have no empirical means to quantify how reliably an AI agent can detect boundary-spanning vulnerabilities. Without a realistic, multi-platform evaluation framework, the security community cannot objectively validate whether state-of-the-art agents possess the cross-contextual reasoning needed to defend the modern, federated identity perimeter.


A key reason this gap is hard to close is what we term the \emph{Correlation Gap}. Unlike traditional text-to-SQL tasks that operate over well-defined, single databases~\cite{yu2018spider}, cross-vendor ISPM requires navigating disjointed systems that lack shared schemas or explicit foreign keys. To succeed, an agent must perform entity resolution across platforms - inferring, for instance, that a human identity in an IdP (\texttt{j.doe@okta.com}) corresponds to the same principal assuming a privileged role in a CSP (\texttt{arn:aws:iam::123:role/JDoe-Dev}). This multi-step reasoning mirrors the data wrangling challenges seen in enterprise-level analytical workflows~\cite{Yao2023_Spider2_Text_to_SQL}. Furthermore, emerging evidence suggests that the primary limitation of frontier LLM-based agents may not be abstract reasoning capability, but rather the absence of structured relational context. Models frequently demonstrate strong security intuition - correctly identifying that a misconfiguration or escalation path exists - while failing to reconstruct the precise supporting entities, join relationships, and cross-platform evidence needed to act on that conclusion. Successful ISPM reasoning therefore depends not only on model intelligence, but on the availability of contextual abstractions that can transform disconnected datasets into a coherent relational graph.

This paper directly addresses this gap. Building on our prior work establishing a foundational benchmark for single-vendor identity visibility tasks~\cite{engelberg2026sola}, we introduce the \textit{Cross-Vendor Sola ISPM Benchmark}, a framework designed to evaluate agentic AI performance across interconnected enterprise ecosystems. The benchmark is deployed over a live, production-grade environment integrating AWS, Okta, Hibob, GCP, MongoDB, GitHub, Azure Active Directory, and Google Workspace, and covers critical cross-vendor use cases including MFA bypass vulnerabilities, dormant account proliferation across federated tenants, and multi-hop privilege escalation vectors.
 
The primary contributions of this work are twofold. First, we introduce a \textbf{cross-vendor ISPM question suite}: 50 data-grounded evaluation tasks deployed over a live, production-grade enterprise environment integrating eight identity platforms. Each task is constructed to require multi-hop entity resolution and cross-system data correlation - for example, tracing a terminated employee record in HiBob through their Okta account to lingering AWS SSO permissions - making the benchmark representative of real-world federated identity investigation workflows. Second, we contribute an \textbf{evidentiary grounding metric framework} designed specifically for cross-vendor ISPM evaluation. Rather than measuring only whether a final answer is correct, the framework assesses the full reasoning chain: structural join reconstruction, multi-schema entity resolution, retrieval utilization quality, and supporting evidence completeness. Together, these two artifacts provide the first standardized means of evaluating whether an agentic system can reliably navigate the Correlation Gap in practice.
The question suite is constructed via an iterative multi-agent pipeline in which a schema-aware generative agent proposes candidate questions and SQL queries, a Critic-Composer loop refines them against expert-defined quality criteria, and a human cybersecurity expert performs final validation. To assess the benchmark, we evaluate the Sola AI Agent under five context configurations - from no injected metadata to full schema, graph, and retrieval context - across three frontier LLMs, yielding a systematic ablation of how structured relational context shapes cross-vendor reasoning fidelity.
 
The remainder of this paper is structured as follows. Section ~\ref{sec:related} describes the related work. Section~\ref{sec:agent} describes the Sola AI Agent architecture, including its schema-grounded execution model and two exploration modes. Section~\ref{sec:benchmark} details the benchmark construction: the integrated data environment, the question elicitation pipeline, dataset composition, and the evaluation metric framework. Section~\ref{sec:experimental_results} presents the experimental setup and results of the ablation study, examining how varying levels of contextual grounding affect agent performance across three frontier LLMs. Section~\ref{sec:conclusion} concludes with a summary of findings and directions for future work.

\section{Related Work}
\label{sec:related}
 
This paper sits at the intersection of two active research areas: text-to-SQL benchmarking for agentic data systems, and AI evaluation for cybersecurity reasoning. We review both and explain where our work departs from them.
 
Spider~1.0~\cite{yu2018spider} established the standard paradigm for evaluating natural-language interfaces to databases, pairing questions with human-authored SQL across a wide range of schemas. Spider~2.0~\cite{Yao2023_Spider2_Text_to_SQL} extended this into realistic enterprise environments, requiring models to handle multi-step workflows and recover from execution errors rather than produce a single correct query in isolation. Our benchmark adopts a similar data-grounded philosophy - every question must be answered from live production data - but introduces a challenge absent from both Spider variants: the absence of shared schemas or explicit foreign keys across vendor boundaries. Where Spider tasks assume a coherent database, cross-vendor ISPM requires the agent to infer join paths across systems that were never designed to interoperate.
 
Several benchmarks evaluate AI reasoning in security-specific contexts. ExCyTIn-Bench~\cite{ExCyTInBench_SecRL} places models in a simulated SOC environment and rewards step-by-step investigative actions, capturing the iterative nature of threat analysis. CyberSOCEval~\cite{CyberSOCEval_CrowdStrike} grounds evaluation in real CrowdStrike~\cite{CrowdStrike} threat reports, testing the ability to extract and synthesize operational intelligence from noisy sources. On the threat intelligence side, CTIBench~\cite{CTIBench} combines CVE/CWE mappings with expert-validated answers, while SEvenLLM-Bench~\cite{SEvenLLM_Bench} introduces bilingual evaluation with hybrid semantic scoring. These benchmarks focus on threat detection and intelligence synthesis; none of them assess whether an agent can query and correlate live identity configuration data across heterogeneous systems.
 
OrgAccess~\cite{OrgAccess_Benchmark_RBAC_Reasoning} is the closest prior work in the IAM space. It evaluates whether models can interpret and apply RBAC rules within synthetic organizational hierarchies. Our benchmark differs in two key respects. First, it operates on production data rather than synthetic structures, grounding every answer in real configurations, access logs, and identity records. Second, it targets operational visibility - enumerating identities, detecting misconfigurations, and tracing cross-system access paths - rather than policy interpretation. The gap we address is not whether a model understands access rules, but whether it can reconstruct the evidence needed to act on them across a fragmented, multi-vendor identity perimeter.

This paper extends our earlier single-vendor ISPM benchmark~\cite{engelberg2026sola}, which established the data-grounded evaluation paradigm used here. We advance it by scaling from one identity platform to eight, redesigning the question suite around cross-vendor multi-hop scenarios, and extending the evaluation framework with structural join metrics and a dual-judge consensus mechanism.
 
Our evaluation framework draws on established LLM-as-a-judge practices, following the rubric-guided scoring approach of G-Eval~\cite{Liu2023_GEVAL} and the retrieval assessment methodology of RAGAS~\cite{Sharma2023_RAGAS}. The use of chain-of-thought prompting~\cite{wei2022chain} in the judge pipeline is consistent with recent findings that structured reasoning improves reliability in complex scoring tasks. We extend these methods with deterministic structural metrics - table and join precision, recall, and F1 against ground truth - that provide a verification layer immune to the semantic ambiguity inherent in purely LLM-based judgment.

\section{Sola AI Agent}
\label{sec:agent}

The Sola AI Agent is a tool-using, data-grounded security AI agent designed to answer complex security questions over enterprise data. While the agent itself is security-domain-generic, this work focuses on its application to Identity Security Posture Management (ISPM), where it is instantiated to address identity inventory and posture visibility questions across enterprise identity and access management environments.

\begin{figure*}[ht]
    \centering
    \includegraphics[width=0.8\textwidth]{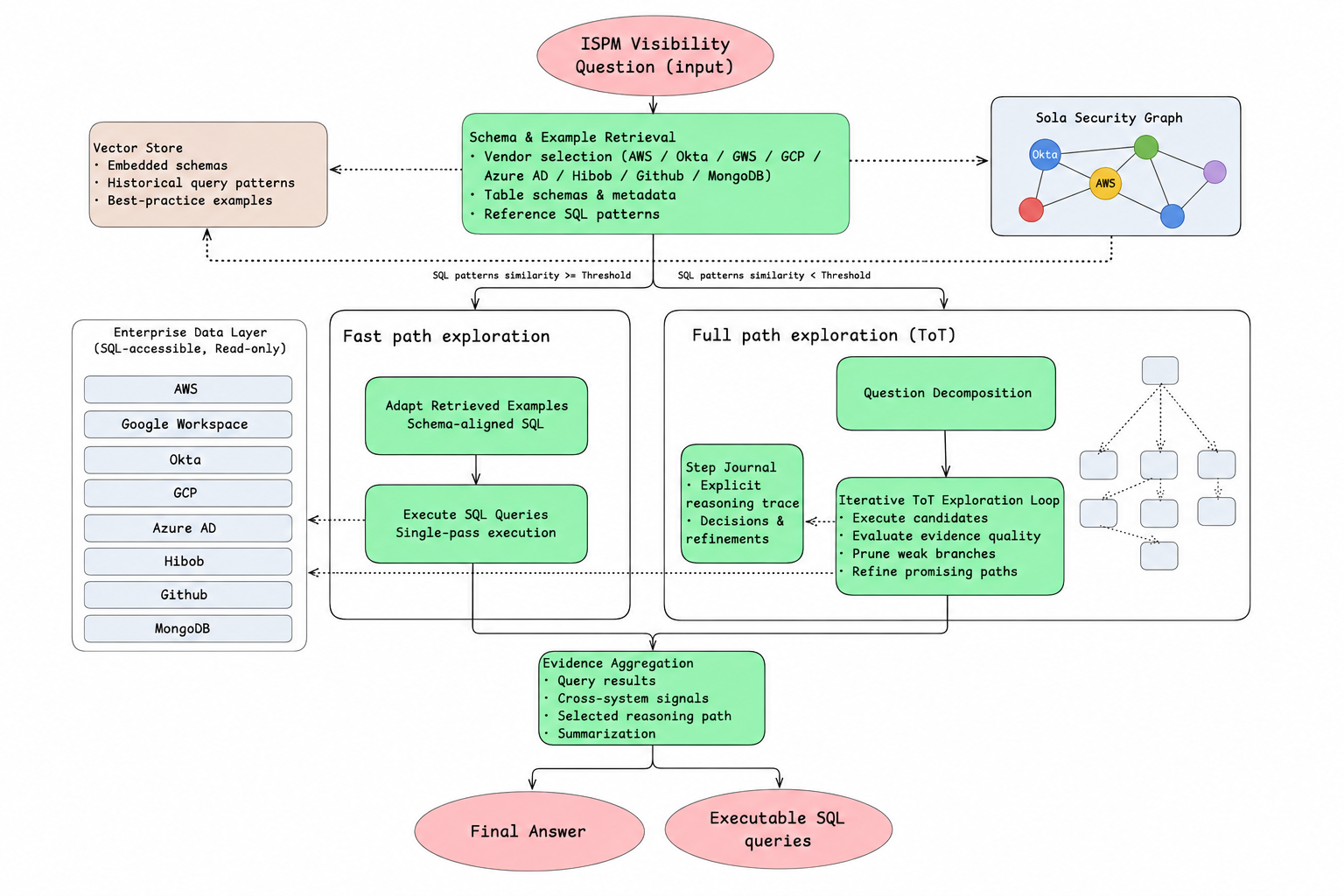}
    \caption{SOLA AI Agent Conceptual Architecture}
    \label{fig:sola-ispm-agent}
\end{figure*}

Given a natural-language ISPM query, the agent translates the request into executable data exploration steps over production IAM, IdP, and SaaS tables, and returns a verifiable, evidence-backed answer. The agent follows a schema-grounded execution model (Figure~\ref{fig:sola-ispm-agent}). For each request, it first identifies the relevant identity platforms and retrieves the corresponding data schemas and reference query patterns. Central to this grounding step is the agent's reliance on Sola's security graph - a unified, cross-vendor mapping of relationships and foreign keys. This graph explicitly defines the logical join paths and relational edges between otherwise disjointed platforms (e.g., linking an Okta user identity to an AWS IAM role). By anchoring its reasoning in this security graph, the agent constrains subsequent actions to valid tables, fields, and cross-system joins, ensuring that all generated queries are executable, data-aligned, and structurally sound. To balance efficiency and robustness, the agent supports two complementary execution modes: fast-path exploration and full-path exploration.

In fast-path exploration, the agent directly adapts retrieved example queries to the target schema and executes them in a single pass. This mode is used when example similarity and schema confidence are high, allowing low-latency enumeration of identity inventories and hygiene conditions without maintaining an explicit reasoning trace. Fast-path execution prioritizes efficiency while remaining fully grounded in underlying data.

In full-path exploration the agent performs an explicit, iterative reasoning process. The question is decomposed into intermediate exploration steps, each associated with a concrete success criterion. The agent executes queries incrementally, validates intermediate results, and records its actions in a structured step journal. This execution pattern is inspired by Tree-of-Thought-style reasoning \cite{yao2023tree}, in which intermediate states are explicitly evaluated and refined, but is uniquely grounded in executable data exploration steps rather than abstract reasoning alone.

In this work, all benchmark evaluations and context ablation experiments are conducted exclusively using the full-path exploration mode. This design choice was made to isolate and measure the impact of contextual grounding on the agent’s explicit reasoning trajectory, intermediate query planning, and evidentiary reconstruction behavior. While fast-path exploration primarily relies on direct adaptation of retrieved examples and therefore benefits less from iterative contextual refinement, full-path execution exposes the agent’s complete reasoning process, making it substantially more suitable for evaluating how schema metadata, graph topology, and retrieved exemplars influence cross-vendor reasoning fidelity and execution stability.

Both execution modes converge on a shared evidence aggregation phase. Query results from all relevant platforms are consolidated and summarized into a final response that includes the natural-language answer, the executed queries, and the supporting evidence, enabling inspection of the agent’s reasoning and decision-making process. When full-path execution is used, the complete step journal is retained, enabling inspection of the agent’s reasoning and decision-making process.

\section{Sola ISPM Visibility Benchmark}
\label{sec:benchmark}

\subsection{Data Source Integration}
\label{subsection:datasource}

The benchmark is grounded in a \textit{live, production-grade enterprise environment} rather than a synthetic or laboratory-generated setup. To authentically replicate the heterogeneous and decentralized nature of contemporary enterprise architectures, this enhanced ecosystem expands significantly beyond the narrower single-domain environments typically used in prior ISPM evaluations, integrating operational identity and configuration data sources from eight distinct platforms. These platforms collectively represent the full technological stack of modern, interconnected organizations: Human Resources Information Systems (HRIS), multiple Identity Providers (IdPs), multi-cloud Cloud Service Providers (CSPs), collaborative productivity suites, developer operations (DevOps) infrastructure, and data persistence layers. This highly integrated, multi-tiered architecture reflects the complex realities of federated enterprise identity ecosystems and establishes the necessary foundation for a rigorous evaluation of cross-vendor identity-security reasoning.

First, Hibob serves as the foundational Human Resources Information System (HRIS), providing the authoritative baseline for human identity lifecycles, including onboarding, departmental affiliations, and offboarding states. Second, an expanded dual-Identity Provider (IdP) layer, implemented using both Okta and Microsoft Azure Active Directory (Azure AD), governs complex authentication flows, federated access policies, and cross-tenant Single Sign-On (SSO) trusts. Third, a multi-cloud execution environment comprising Amazon Web Services (AWS) and Google Cloud Platform (GCP) hosts the organization's operational workloads, capturing intricate Identity and Access Management (IAM) roles, cross-account permissions, and non-human service principals. Fourth, Google Workspace (GWS) continues to represent the collaborative and resource-governance layer, managing workforce communications, document sharing, and OAuth application integrations. Finally, GitHub provides visibility into the critical infrastructure and developer operations (DevOps) tier tracking source control permissions and developer tokens while MongoDB operates as the data persistence layer, exposing sensitive asset repositories subject to granular, database-level access controls.

Integrating these eight operational platforms creates a unified yet highly complex identity environment in which all benchmark questions must be resolved. Rather than querying isolated data silos, the agent must perform multi-hop entity resolution and cross-system access tracing, such as mapping a terminated employee in HiBob to lingering active sessions in Microsoft Azure Active Directory, or linking an identity in Okta to privileged deployment tokens in GitHub and Amazon Web Services. Importantly, all benchmark tasks are strictly data-grounded: every answer, including complex inter-system joins and federated trust validations, must be derived solely from the configurations, relationships, logs, and policies contained within the integrated environment, without reliance on external assumptions or prior knowledge.

\subsection{ISPM Question Elicitation}

\begin{figure}[ht!]
    \centering
    \includegraphics[width=\linewidth]{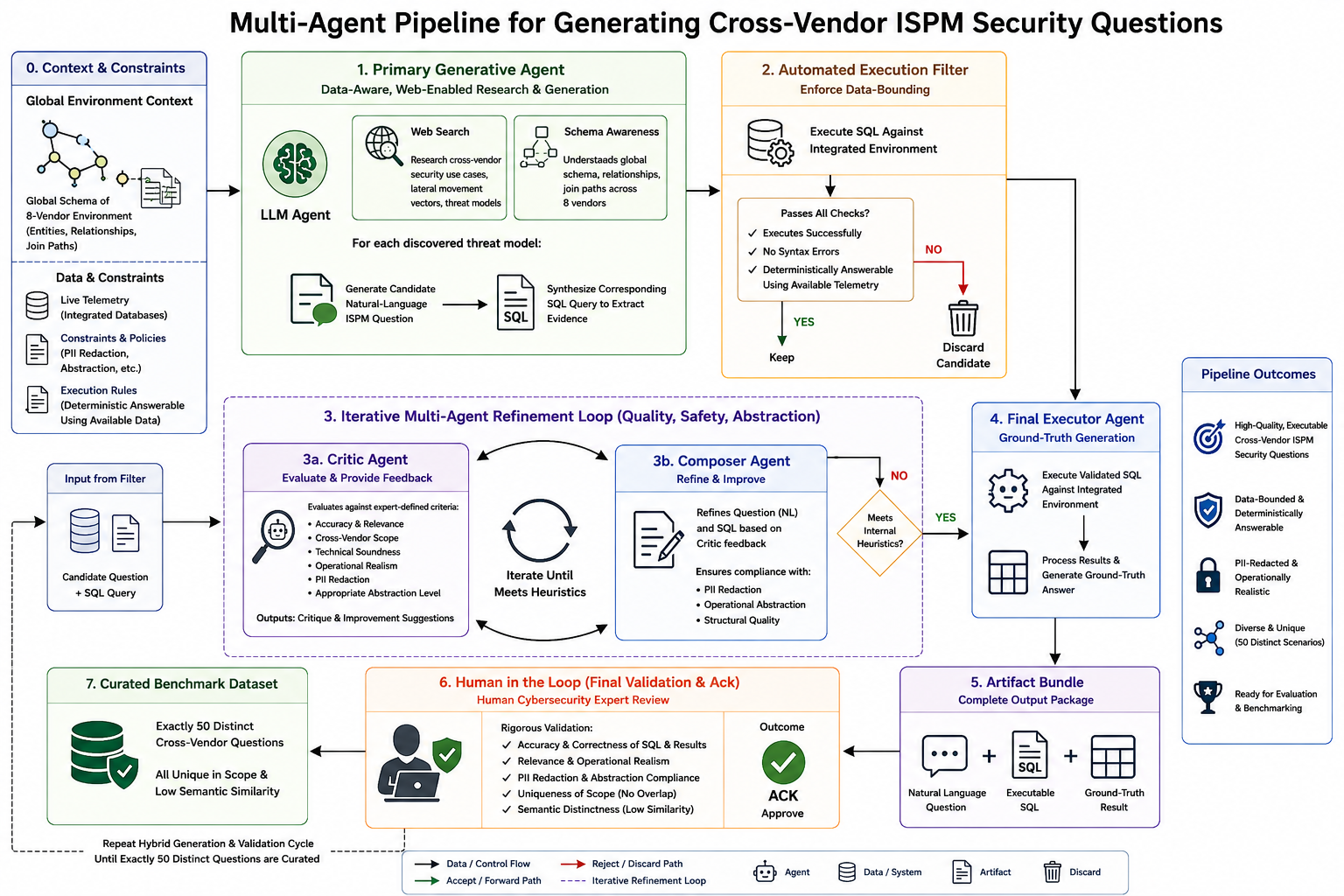}
    \caption{Iterative, Multi-Agent Question Generation and Validation Pipeline.}
    \label{fig:question-generation-pipeline}
\end{figure}

To systematically construct realistic cross-vendor ISPM scenarios, we designed the iterative multi-agent generation pipeline illustrated in Figure~\ref{fig:question-generation-pipeline}. The pipeline was seeded with a curated set of high-value identity security use cases selected by cybersecurity domain experts, ensuring that the generated benchmark reflects operationally relevant attack paths, lifecycle failures, privilege escalation scenarios, and cross-vendor identity risks commonly encountered in enterprise environments. These expert-defined security themes guided the subsequent automated generation process.

The generation process begins with a primary generative agent that is granted direct visibility into the global schema of the integrated environment, enabling awareness of cross-vendor relationships and potential join paths across participating platforms. To improve operational realism, the agent is additionally equipped with web-search capabilities and prompted to autonomously research realistic attack paths, identity misconfigurations, and lateral movement scenarios commonly observed in enterprise environments. For each discovered threat model, the agent synthesizes both a candidate natural-language ISPM question and a corresponding SQL query intended to extract supporting evidence from the integrated environment.

To enforce strict data grounding, all generated candidates are passed through an automated execution filter. At this stage, candidate questions are discarded if their paired SQL queries fail to execute, produce syntax errors, or cannot be deterministically answered using the available telemetry and relationships present within the environment. Only candidates that successfully satisfy these executability constraints are forwarded to the subsequent refinement stages.

Validated candidates are then routed into an iterative multi-agent refinement loop composed of a \textit{Critic} agent and a \textit{Composer} agent. The Critic evaluates each candidate against expert-defined criteria including technical correctness, operational realism, cross-vendor relevance, abstraction quality, structural soundness, and PII compliance. Based on this feedback, the Composer iteratively refines both the natural-language question and its associated SQL query. This refinement cycle repeats until the candidate satisfies all internal quality heuristics enforced by the pipeline.

Once validated, a final executor agent executes the refined SQL query against the integrated environment to generate the definitive ground-truth answer. The resulting artifact bundle, consisting of the natural-language question, executable SQL query, and ground-truth output, is subsequently subjected to a final review by a human cybersecurity expert. This validation stage verifies execution correctness, operational realism, semantic uniqueness, and compliance with abstraction and privacy requirements.

The complete hybrid generation and validation cycle was repeated iteratively until a curated benchmark of 50 operationally distinct cross-vendor ISPM scenarios was obtained. To preserve benchmark diversity and reduce redundancy, we additionally enforced low semantic similarity and non-overlapping operational scope across all curated questions.

\subsection{Cross-Vendor Dataset Composition and Complexity}

To facilitate robust cross-vendor analysis, the benchmark encompasses a diverse ecosystem of identity and service providers, ensuring evaluated agents must reason over heterogeneous schemas. As detailed in Table \ref{tab:vendor_distribution}, the dataset is anchored by major Identity Providers (IdPs) such as Okta (27.2\%) and Azure AD (5.3\%), while integrating critical cloud and developer infrastructure (AWS at 12.3\%, GCP at 3.5\%, GitHub at 14.0\%, MongoDB Atlas at 2.6\%), collaborative productivity suites like Google Workspace (16.7\%), and specialized SaaS applications including HiBob. This distribution reflects the tri-layer identity architecture typical of modern cloud-native enterprises, requiring agents to reconcile identity definitions across boundaries. 

\begin{table}[htbp]
\centering
\begin{tabular}{lc}
\toprule
\textbf{Platform / Vendor} & \textbf{Number of Questions} \\
\midrule
Okta & 31 \\
HiBob & 19 \\
Google Workspace & 19 \\
GitHub & 16 \\
AWS & 14 \\
Azure AD & 6 \\
GCP & 4 \\
MongoDB Atlas & 3 \\
Hibob & 2 \\
\bottomrule
\end{tabular}
\caption{Distribution of benchmark questions across identity and service providers.}
\label{tab:vendor_distribution}
\end{table}

Beyond schema diversity, we evaluate task difficulty using a multi-dimensional complexity framework, detailed in Table \ref{tab:complexity_stats}. First, we apply the syntactic complexity criteria from Spider 2.0 \cite{Yao2023_Spider2_Text_to_SQL}, which categorizes queries based on SQL token length. During the curation of the ground-truth dataset, we deliberately optimized the reference SQL queries for execution efficiency and conciseness. Consequently, a substantial majority (74.0\%) of our queries classify as Easy under this token-based metric. While traditional text-to-SQL benchmarks often equate difficulty with query verbosity, such length-based assessments do not adequately reflect the structural and relational challenges inherent to the ISPM domain, where the logical connections between disparate tables are highly non-trivial. 

To accurately capture these unique operational reasoning demands, we introduce two structure-aware metrics. We measure \textit{Structural Complexity} by the absolute number of \texttt{JOIN} operations required, capturing the breadth of tables an agent must connect. To quantify the depth of relational reasoning, we introduce \textit{Relational Depth (Maximum Join Path Length)}, which models the query as a schema graph and calculates the longest sequence of dependent joins (hops) the agent must sequentially traverse. This distinction is critical: a query connecting one central table to two independent dimension tables requires multiple joins but only a single hop. 
\begin{table}[ht!]
\centering
\begin{tabular}{lc}
\toprule
\textbf{Complexity Metric} & \textbf{Distribution} \\
\midrule
\multicolumn{2}{l}{\textbf{Syntactic Complexity (Spider 2.0 Token Length)}} \\
- Easy ($\text{tokens} < 80$) & 37 (74.0\%) \\
- Medium ($80 \le \text{tokens} < 160$) & 12 (24.0\%) \\
- Hard ($\text{tokens} \ge 160$) & 1 (2.0\%) \\
\midrule
\multicolumn{2}{l}{\textbf{Structural Complexity (Required JOINs)}} \\
- Easy ($1$ joins) & 22 (44.0\%) \\
- Medium ($2$ or $3$ joins) & 26 (52.0\%) \\
- Hard ($\ge 3$ joins) & 2 (4.0\%) \\
\midrule
\multicolumn{2}{l}{\textbf{Relational Depth (Maximum Join Path Length / Hops)}} \\
- Easy ($\le 1$ hop) & 25 (50.0\%) \\
- Medium ($2$ hops) & 22 (44.0\%) \\
- Hard ($\ge 3$ hops) & 3 (6.0\%) \\
\bottomrule
\end{tabular}
\caption{Distribution of benchmark questions evaluated by syntactic SQL length, absolute required JOINs, and Relational Depth (Hops).}
\label{tab:complexity_stats}
\end{table}

\subsection{Evaluation Framework}
\label{sub:eval_framework}

\begin{figure}[ht]
    \centering
    \includegraphics[width=0.8\textwidth]{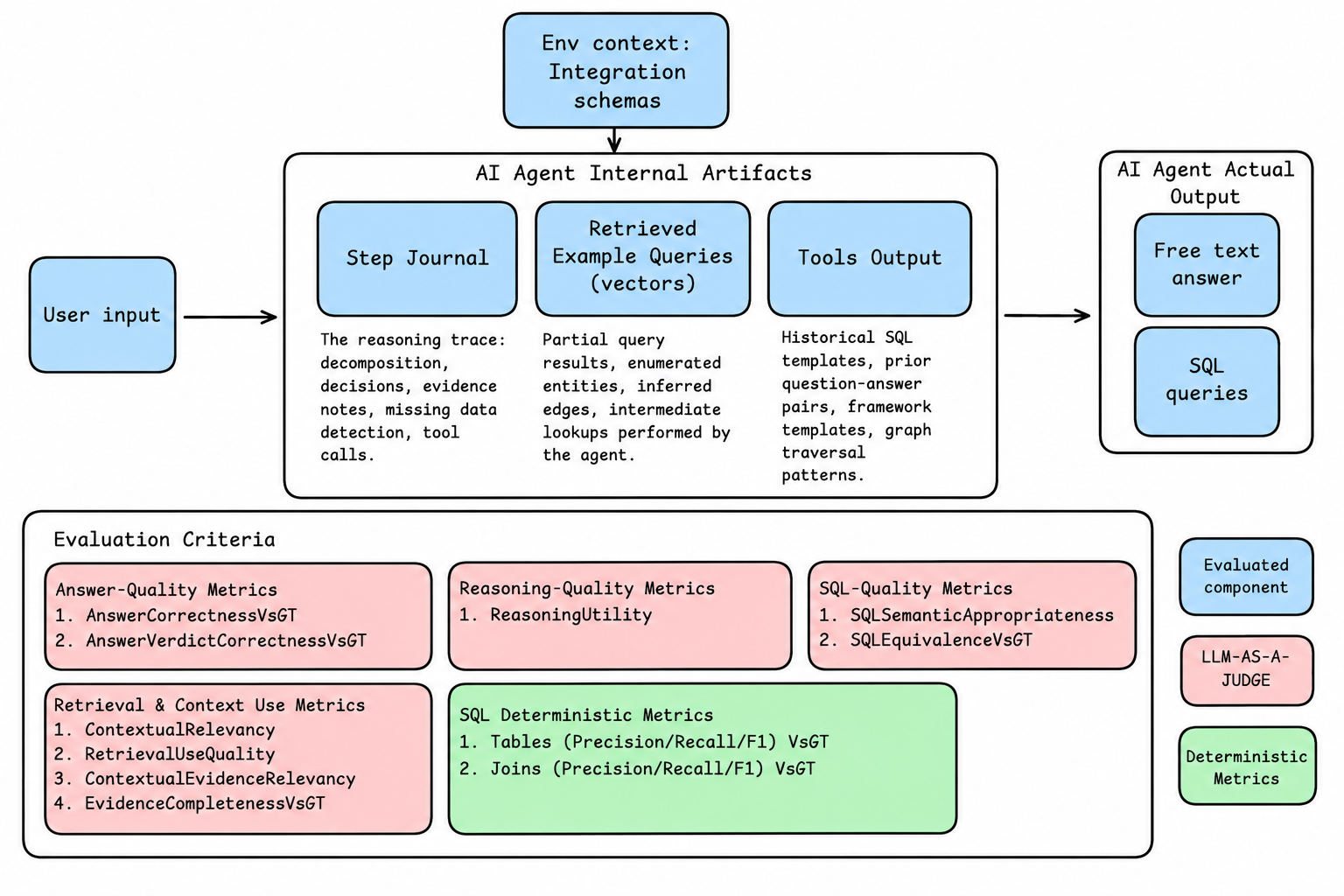}
    \caption{Sola ISPM Visibility Evaluation Framework}
    \label{fig:framework}
\end{figure}

Modern agentic AI systems for enterprise data question answering share a common architectural pattern composed of several interconnected components: a natural-language interpretation layer that parses the user query, a retrieval layer that locates relevant schema elements, examples, or historical traces, a reasoning engine that constructs a multi-step plan, a tool-execution layer responsible for generating and running SQL or API calls, and a synthesis layer that integrates intermediate results into a final natural-language answer. Although implementations differ, these components form the backbone of common agentic systems, and each introduces opportunities for both correct and faulty behavior. A comprehensive evaluation must therefore examine not only the final answer, but also the internal reasoning and the evidence used at each stage of the computation. To evaluate these components in a controlled and reproducible manner, We introduce a four-stage process.

\paragraph{Step 1: Executing All Questions Through the Sola Agent.}
After curating the benchmark question set, every question is executed end-to-end through the Sola agent. This execution yields a complete agentic reasoning trace, including the natural-language interpretation of the query, the reasoning steps taken, the generated SQL queries, and the results they yield. This approach mirrors the data-grounded evaluation paradigm introduced in data-aware text-to-SQL benchmarks such as Spider~2.0 \cite{Yao2023_Spider2_Text_to_SQL}, ensuring that both the reasoning path and the outputs are fully observable and assessable.

\paragraph{Step 2: Collection of the Evidence Bundle.}
For each question, we collect a structured set of evaluation artifacts that represent the behavior of different components of the agentic pipeline: \textit{Input}: the original ISPM question posed to the agent. \textit{Output (Natural Language)}: the final answer produced by the synthesis layer. \textit{Generated SQL}: the queries constructed by the tool-execution layer to retrieve the final answer. \textit{Retrieved Example Queries}: embeddings and examples surfaced by the retrieval layer, representing schema hints, patterns, and prior examples available to the agent. \textit{Step Journal and Tools Output}: a detailed, step-by-step record of the agent’s internal reasoning process, reflecting plan construction, schema interpretation, query formulation, synthesis logic, and collected evidence from executing SQL queries against the environment. Each artifact provides insight into a different subsystem of the agent. Together, they form the evidence bundle used for expert evaluation and LLM-based judgment.


\paragraph{Step 3: LLM-as-Judge Evaluation.} To comprehensively evaluate the complex cognitive processes required for cross-vendor ISPM, we deploy an expanded LLM-as-a-Judge pipeline, applying structured chain-of-thought (CoT) \cite{wei2022chain} scoring across a three-level rubric: 0 (``Does Not Meet Criterion''), 0.5 (``Partially Meets Criterion''), and 1 (``Fully Meets Criterion''). This approach follows the rubric-guided evaluation paradigm introduced by G-Eval \cite{Liu2023_GEVAL}, where LLM judges reason explicitly before issuing a score. Each benchmark instance is evaluated by a panel of two frontier models: Anthropic Claude Sonnet 4.6 \cite{Anthropic2026_Sonnet46} and OpenAI GPT-4.1 \cite{OpenAI2026_GPT41}, selected for their complementary strengths. Sonnet 4.6 provides strong long-context and schema-aware reasoning, making it highly effective for evaluating deep step journals and cross-vendor multi-hop logic. Conversely, GPT-4.1 provides semantically precise judgments and stable Q\&A performance under moderate token constraints. 
We implement a multi-sample majority 
voting mechanism to produce a stable, bias-corrected consensus. The final 
score $S$ for a given metric is the statistical mode of the concatenated pool 
of all individual judge evaluations:

$$S = \operatorname{mode} \left( \bigoplus_{j=1}^{J} \bigoplus_{t=1}^{T} s_{j,t} \right)$$

Where $J$ represents the total number of distinct LLM judge models in the evaluation panel, $T$ represents the total number of independent execution traces generated by the agent for each benchmark instance, and $s_{j,t} \in \{0, 0.5, 1\}$ represents the discrete rubric score assigned by judge $j$ for execution trace $t$. The sequence concatenation operator $\bigoplus$ aggregates the multi-dimensional array of individual judgments into a single, flattened pool, from which the $\operatorname{mode}$ extracts the strict majority vote consensus. 

In our specific implementation, we employ $J=2$ frontier models and run $T=5$ independent execution traces per query. This flattens the evaluations into a combined sequence of exactly 10 outcomes. This consensus model effectively smooths out individual judge variances. For instance, if one judge assigns scores of $\{0, 0, 0, 1, 0.5\}$ and the other assigns $\{1, 1, 1, 1, 0.5\}$, the overall majority vote successfully evaluates to $1$.

This dual-judge panel evaluates four primary components: (1) the user's question, (2) the agent's internal artifacts (step journal, retrieved examples, tool outputs), (3) the agent's actual outputs (free-text and SQL), and (4) the expert-validated ground truth. Evaluations are conducted across a structured, nine-metric suite organized into four established evaluation criteria groups: \textit{Retrieval \& Context Use, Reasoning-Quality, SQL-Quality,} and \textit{Answer-Quality}. 

To maintain strict statistical hygiene, the framework designates metrics as non-applicable (N/A) - logging explicit skip reasons - whenever a specific execution instance renders them undefined (e.g., when ground-truth queries yield empty entity sets). This structural safeguard prevents the artificial inflation or deflation of aggregate performance scores, ensuring equitable comparisons across differing agent architectures.

To operationalize this framework, the constituent elements of our evaluation suite - comprising nine LLM-evaluated metrics and three derived aggregations - are categorized and defined within their respective evaluation groups as follows:

\begin{itemize}

\item \textbf{Answer-Quality Metrics:} 
    \begin{itemize}
        \item \texttt{AnswerCorrectnessVsGT} ($A.Corr$): Assesses semantic equivalence with the ground truth across entities, counts, and security posture classifications \cite{Sharma2023_RAGAS}. 
        \item \texttt{AnswerVerdictCorrectnessVsGT} ($A.Ver$): Provides a binary verification of whether the primary yes/no verdict matches the ground truth, decoupling core correctness from weaker supporting details. Conceptually, \texttt{AnswerVerdictCorrectnessVsGT} serves as an upper-bound simplification of \texttt{AnswerCorrectnessVsGT}, isolating whether the model reached the correct high-level security conclusion independently from the completeness and precision of its supporting evidence.
        
        \vspace{0.5em} 
        \noindent\textit{Derived Performance Aggregations:} To summarize overall system efficacy, we compute three deterministic metrics directly from the \texttt{AnswerCorrectnessVsGT} score distribution:
        \begin{itemize}
            \item \texttt{SuccessRate} ($SR$): Calculates strict accuracy by measuring the proportion of instances where \texttt{AnswerCorrectnessVsGT} scored exactly 1. 
            \item \texttt{SuccessRateAtLeast0.5} ($SR\ge0.5$): Computes partial-to-full success by aggregating instances that scored 0.5 or 1.
            \item \texttt{FailureRate} ($FR$): Measures complete inaccuracies by isolating instances that scored exactly 0.
        \end{itemize}
    \end{itemize}

    \item \textbf{Retrieval \& Context Use Metrics:} 
    \begin{itemize}
        \item \texttt{ContextualRelevancy} ($C.Rel$): Evaluates the alignment of the final retrieved candidates vectors consumed by the agent \cite{Sharma2023_RAGAS}. 
        \item \texttt{RetrievalUseQuality} ($R.Use$): Assesses whether the agent effectively utilized these retrieved candidates when formulating its SQL, specifically penalizing irrelevant template reuse and missed opportunities. 
        \item \texttt{ContextualEvidenceRelevancy} ($C.Ev$): Conducts a strict entity-by-entity comparison of the agent's SQL results against the ground truth's SQL results. 
        \item \texttt{EvidenceCompletenessVsGT} ($E.Comp$): Verifies whether the broader supporting facts within the answer - such as named users, account statuses, and departments - properly overlap with the curated ground truth lists.
    \end{itemize}
    
    \item \textbf{Reasoning-Quality Metrics:} 
    \begin{itemize}
        \item \texttt{ReasoningUtility} ($R.Ut$): Evaluates whether the agent is on the correct path, determining if its table choices and sub-questions plausibly progress toward a correct resolution \cite{DeepEval_StepEfficiency}.
    \end{itemize}
    
    \item \textbf{SQL-Quality Metrics:} 
    \begin{itemize}
        \item \texttt{SQLSemanticAppropriateness} ($S.Sem$): Functions as a schema-aware generic check to ensure the generated SQL is a sensible adaptation for the question \cite{Yao2023_Spider2_Text_to_SQL}. 
        \item \texttt{SQLEquivalenceVsGT} ($S.Eq$): Compares the agent's generated SQL to verify logical equivalence with the curated ground truth's SQL \cite{Min2023_Benchmarking_Text_to_SQL_LLMs}.
    \end{itemize}
\end{itemize}

\paragraph{Step 4: Deterministic Structural Evaluation.} While the LLM-as-a-Judge pipeline provides nuanced semantic assessments, cross-vendor ISPM fundamentally depends on the agent's ability to accurately link disparate datasets across vendor boundaries. To explicitly quantify this structural understanding, we introduce a dedicated suite of deterministic metrics that serve as a rigorous, structural complement to our LLM-based SQL-Quality metrics. The underlying logic dictates that while specific filtering conditions or intermediate SQL syntax may vary between a valid agent output and the ground truth, the core tables utilized and the relational edges (JOINs) connecting them must align to reach a correct cross-vendor resolution.

To systematically compute these metrics, we utilized an LLM specifically prompted to parse and extract the utilized tables and JOIN conditions from both the agent's generated SQL and the ground-truth SQL. To guarantee absolute extraction fidelity and prevent LLM hallucinations, a strict programmatic validation layer was implemented. This layer verifies that any extracted table or JOIN explicitly exists within the raw SQL string; if a hallucination is detected, the extraction is iteratively re-prompted. Once reliably extracted, the agent's tables and joins are compared directly against the ground-truth sets. By treating the existence of each table or join as a binary classification (1 or 0), we compute standard precision, recall, and F1-scores.

The deterministic structural metrics are defined as follows:

\begin{itemize}
    \item \textbf{Tables (Precision/Recall/F1) VsGT} ($P/R/F1$): Deterministically computes the exact match of database tables queried by the agent versus the ground truth. Precision penalizes inefficient over-fetching (noise) and reliance on unnecessary datasets, while recall penalizes the omission of critical data sources. 
    
    \item \textbf{Joins (Precision/Recall/F1) VsGT} ($P/R/F1$): Deterministically evaluates the specific graph edges (joins) constructed between tables, assessing whether the agent successfully reproduced the correct cross-vendor entity resolution paths.
\end{itemize}

\section{Ablation Study}
\label{sec:experimental_results}

This section presents the empirical evaluation of the Sola agent architecture. Moving beyond isolated, domain-specific performance checks, we conduct a systematic ablation study to isolate the precise impact of varying contextual abstraction levels across several leading frontier LLMs via our derived metrics from \ref{sub:eval_framework}, allowing for a transparent assessment of absolute system reliability.


\subsection{Experimental Setup}
\label{subsec:experimental_setup}

To rigorously evaluate the Sola framework, we experimented the agent across three state-of-the-art foundation models: \textbf{Anthropic Claude 4.6 Sonnet}, \textbf{Anthropic Claude 4.8 Opus}, \textbf{GPT 5.5}, and \textbf{Google Gemini 3.1 Pro}. Each experiment was subjected to five distinct context configurations to map how varying levels of metadata abstraction alter reasoning and code generation quality. These configurations are defined as follows:

\begin{itemize}
    \item \textbf{No Context}: The agent is completely starved of pre-injected metadata, schema assets, or semantic assistance. However, it retains raw access to the interactive execution environment. To construct valid multi-hop queries, it must engage in blind, autonomous environment discovery, iteratively issuing trial queries, capturing execution feedback, and attempting to infer table paths and field mappings on the fly.
    \item \textbf{Schema Only}: The agent is provided only with the raw data definitions and schemas of the relevant vendor database tables, providing a basic relational baseline.
    \item \textbf{Schema + Graph}: The agent maintains access to the relevant vendor database schemas, identical to the \textit{Schema Only} configuration, but is additionally supplemented by Sola's specialized Security Graph context, which explicitly maps cross-vendor entity resolution edges (JOIN paths) without providing candidate examples.
    \item \textbf{Schema + Examples}: The agent receives table schemas and a set of retrieved few-shot candidate examples containing relevant text-to-SQL pairs, but lacks the unified topological graph context. These candidates retrieved from Sola proprietary security insights knowledge base.
    \item \textbf{Full Context}: The complete Sola production pipeline, providing the agent with table schemas, retrieved candidate examples, and full Security Graph context.
\end{itemize}

The cross-product of these four models and five context settings establishes twenty unique experimental combinations. Every configuration was comprehensively evaluated against our expert-curated ground truth utilizing the multi-sample consensus pipeline and the nine-metric, three-aggregation evaluation framework detailed in Section \ref{sub:eval_framework}. 

Crucially, because the underlying Sola agent architecture supports interactive runtime loops, the agent can execute code and process live database responses across all five settings. A summary of the contextual features explicitly pre-injected per configuration is provided in Table \ref{tab:context_configurations}.

\begin{table}[ht]
    \centering
    \small
    \begin{tabular}{lccc}
        \hline
        \textbf{Context Configuration} & \textbf{Table Schemas} & \textbf{Security Graph} & \textbf{Candidate Examples} \\
        \hline
        No Context & \texttimes & \texttimes & \texttimes \\
        Schema Only & \checkmark & \texttimes & \texttimes \\
        Schema + Graph & \checkmark & \checkmark & \texttimes \\
        Schema + Examples & \checkmark & \texttimes & \checkmark \\
        Full Context & \checkmark & \checkmark & \checkmark \\
        \hline
    \end{tabular}
    \caption{Feature Matrix of the Five Context Ablation Configurations.}
    \label{tab:context_configurations}
\end{table}

\subsection{Experimental Results}
\label{subsec:overall_results}

Tables \ref{tab:ablation_gpt}, \ref{tab:ablation_gemini}, \ref{tab:ablation_sonnet}, and \ref{tab:ablation_opus} report the macro quantitative evaluation across all context configurations for GPT 5.5, Gemini 3.1 Pro, Claude 4.6 Sonnet, and Claude 4.8 Opus, respectively. Across each model matrix, the metrics are segmented into our three primary LLM-as-a-Judge dimensions - \textit{Retrieval \& Context Use}, \textit{Reasoning}, and \textit{SQL Quality} - alongside final \textit{Answer Quality} tracks and deterministic evaluations. To provide a consolidated view of peak operational performance within this cross-vendor evaluation, Table \ref{tab:summary_answer_correctness} isolates the optimal context configuration for each frontier model specifically targeting the core \texttt{AnswerCorrectnessVsGT} metric.

\begin{table}[ht!]
    \centering
    \small
    \renewcommand{\arraystretch}{1.2}
    \begin{tabular}{lcccc}
        \hline
        \textbf{Context Configuration} & \textbf{GPT 5.5} & \textbf{Gemini 3.1 Pro} & \textbf{Claude 4.6 Sonnet} & \textbf{Claude 4.8 Opus} \\
        \hline
        No Context         & 0.54 & 0.58 & 0.53 & 0.58 \\
        Schema Only        & 0.62 & 0.65 & 0.64 & 0.68 \\
        Schema + Graph     & 0.66 & 0.67 & 0.7 & 0.73 \\
        Schema + Examples  & 0.67 & 0.72 & 0.72 & 0.76 \\
        Full Context       & \textbf{0.69} & \textbf{0.74} & \textbf{0.76} & \textbf{0.78} \\
        \hline
    \end{tabular}
    \caption{Consolidated \texttt{AnswerCorrectnessVsGT} performance matrix across all models and context configurations.}
    \label{tab:summary_answer_correctness}
\end{table}

\subsubsection{Results Overview}
Our empirical results definitively highlight that the \textbf{Full Context} configuration under the \textbf{Claude 4.8 Opus} engine yields the highest absolute cross-vendor performance, maximizing semantic \texttt{AnswerCorrectnessVsGT} (0.78) and achieving the lowest overall \texttt{FailureRate} (4.0\%). Across the evaluated cohort, performance generally scales with the richness of the injected context layer. When isolating the foundation models themselves, Claude 4.8 Opus systematically outperforms GPT 5.5, Claude 4.6 Sonnet, and Gemini 3.1 Pro across \texttt{AnswerCorrectnessVsGT} at almost every configuration tier. Notably, GPT 5.5 achieved a peak \texttt{AnswerCorrectnessVsGT} score of 0.69 under Full Context, demonstrating consistent gains from richer contextual grounding while still remaining below the strongest frontier reasoning models.

For all evaluated models, combining all structural context layers produced the strongest overall performance under the Full Context configuration.
The \texttt{JoinsF1VSGroundTruth} metric exhibits a clear upward trajectory as context complexity scales, consistently peaking whenever graph knowledge is injected - either in the isolated \textbf{Schema + Graph} or the consolidated \textbf{Full Context} configurations. For instance, Gemini's join fidelity surges from a baseline of 0.55 to a peak of 0.68, and Sonnet spikes to 0.66. Similarly, GPT 5.5 demonstrates a measurable improvement in structural routing, with its join F1 score increasing from 0.45 under Schema Only to 0.51 under Full Context, while recall notably rises from 0.61 to 0.73 once graph-aware context is incorporated. This empirical trend strongly implies that the engines actively leverage the relational topologies explicitly mapped within the graph. By internalizing these structural nodes, the models develop a fundamentally more accurate schema-linking capability, allowing them to correctly identify and execute complex relational joins across disparate vendor databases even when overall natural language generation fluctuates.

\begin{table*}[ht!]
    \centering
    \scriptsize
    \setlength{\tabcolsep}{2.5pt}
    \begin{tabular}{l | cccc | c | cc | ccccc | ccc | ccc}
        \hline
        \textbf{Context} & \multicolumn{4}{c|}{\textbf{Retrieval \& Context Use}} & \textbf{Reasoning} & \multicolumn{2}{c|}{\textbf{SQL Quality}} & \multicolumn{5}{c|}{\textbf{Answer Quality}} & \multicolumn{3}{c|}{\textbf{Structural Tables}} & \multicolumn{3}{c}{\textbf{Structural Joins}} \\
        \textbf{Configuration} & \textbf{C.Rel} & \textbf{R.Use} & \textbf{C.Ev} & \textbf{E.Comp} & \textbf{R.Ut} & \textbf{S.Sem} & \textbf{S.Eq} & \textbf{A.Corr} & \textbf{A.Ver} & \textbf{SR} & \textbf{SR$\ge$0.5} & \textbf{FR} & \textbf{P} & \textbf{R} & \textbf{F1} & \textbf{P} & \textbf{R} & \textbf{F1} \\
        \hline
        \textbf{No Context} & -- & 0.66 & 0.62 & 0.58 & 0.95 & 0.95 & 0.62 & 0.56 & 0.78 & 44.0\% & 68.0\% & 32.0\% & 0.36 & 0.78 & 0.47 & 0.43 & 0.58 & 0.43 \\
        \textbf{Schema Only} & -- & 0.99 & 0.66 & 0.63 & 0.99 & 0.99 & 0.74 & 0.64 & 0.88 & 48.0\% & 80.0\% & 20.0\% & 0.56 & 0.81 & 0.64 & 0.37 & 0.57 & 0.39 \\
        \textbf{Schema + Graph} & -- & 0.99 & 0.73 & 0.69 & 0.99 & 0.99 & \textbf{0.81} & 0.70 & 0.88 & 58.1\% & 81.4\% & 18.6\% & 0.60 & 0.90 & 0.69 & 0.42 & 0.73 & 0.46 \\
        \textbf{Schema + Examples} & -- & 1.00 & 0.77 & 0.67 & 0.99 & 0.99 & 0.77 & 0.67 & 0.90 & 48.0\% & 86.0\% & 14.0\% & 0.63 & 0.88 & 0.70 & \textbf{0.48} & \textbf{0.79} & \textbf{0.53} \\
        \textbf{Full Context} & -- & 1.00 & \textbf{0.81} & \textbf{0.77} & 0.99 & \textbf{1.00} & 0.77 & \textbf{0.74} & \textbf{0.91} & \textbf{60.5\%} & \textbf{88.4\%} & \textbf{11.6\%} & \textbf{0.63} & \textbf{0.90} & 0.70 & 0.48 & 0.79 & 0.52 \\
        \hline
    \end{tabular}
    \caption{Comprehensive Context Ablation Matrix for \textbf{GPT 5.5}.}
    \label{tab:ablation_gpt}
\end{table*}

\begin{table*}[ht!]
    \centering
    \scriptsize
    \setlength{\tabcolsep}{2.5pt}
    \begin{tabular}{l | cccc | c | cc | ccccc | ccc | ccc}
        \hline
        \textbf{Context} & \multicolumn{4}{c|}{\textbf{Retrieval \& Context Use}} & \textbf{Reasoning} & \multicolumn{2}{c|}{\textbf{SQL Quality}} & \multicolumn{5}{c|}{\textbf{Answer Quality}} & \multicolumn{3}{c|}{\textbf{Structural Tables}} & \multicolumn{3}{c}{\textbf{Structural Joins}} \\
        \textbf{Configuration} & \textbf{C.Rel} & \textbf{R.Use} & \textbf{C.Ev} & \textbf{E.Comp} & \textbf{R.Ut} & \textbf{S.Sem} & \textbf{S.Eq} & \textbf{A.Corr} & \textbf{A.Ver} & \textbf{SR} & \textbf{SR$\ge$0.5} & \textbf{FR} & \textbf{P} & \textbf{R} & \textbf{F1} & \textbf{P} & \textbf{R} & \textbf{F1} \\
        \hline
        \textbf{No Context} & -- & 0.75 & 0.56 & 0.58 & 0.94 & 0.96 & 0.61 & 0.58 & 0.86 & 46.0\% & 70.0\% & 30.0\% & 0.39 & 0.81 & 0.51 & 0.59 & 0.56 & 0.56 \\
        \textbf{Schema Only} & -- & 0.97 & 0.59 & 0.64 & 0.96 & 0.96 & 0.69 & 0.65 & 0.92 & 50.0\% & 80.0\% & 20.0\% & 0.68 & 0.89 & 0.74 & 0.50 & 0.67 & 0.52 \\
        \textbf{Schema + Graph} & -- & 1.00 & 0.70 & 0.66 & 1.00 & 1.00 & 0.69 & 0.68 & 0.94 & 50.0\% & 86.0\% & 14.0\% & 0.75 & 0.89 & 0.79 & 0.60 & 0.76 & 0.61 \\
        \textbf{Schema + Examples} & -- & 1.00 & \textbf{0.74} & 0.73 & 1.00 & 1.00 & 0.68 & \textbf{0.74} & 0.96 & 54.0\% & \textbf{94.0\%} & \textbf{6.0\%} & 0.72 & 0.90 & 0.76 & 0.59 & 0.74 & 0.60 \\
        \textbf{Full Context} & -- & 1.00 & 0.72 & \textbf{0.77} & 0.98 & 0.99 & \textbf{0.72} & 0.73 & \textbf{0.98} & \textbf{56.0\%} & 90.0\% & 10.0\% & \textbf{0.79} & \textbf{0.91} & \textbf{0.82} & \textbf{0.65} & 0.76 & \textbf{0.66} \\
        \hline
    \end{tabular}
    \caption{Comprehensive Context Ablation Matrix for \textbf{Gemini 3.1 Pro}.}
    \label{tab:ablation_gemini}
\end{table*}

\begin{table*}[ht!]
    \centering
    \scriptsize
    \setlength{\tabcolsep}{2.5pt}
    \begin{tabular}{l | cccc | c | cc | ccccc | ccc | ccc}
        \hline
        \textbf{Context} & \multicolumn{4}{c|}{\textbf{Retrieval \& Context Use}} & \textbf{Reasoning} & \multicolumn{2}{c|}{\textbf{SQL Quality}} & \multicolumn{5}{c|}{\textbf{Answer Quality}} & \multicolumn{3}{c|}{\textbf{Structural Tables}} & \multicolumn{3}{c}{\textbf{Structural Joins}} \\
        \textbf{Configuration} & \textbf{C.Rel} & \textbf{R.Use} & \textbf{C.Ev} & \textbf{E.Comp} & \textbf{R.Ut} & \textbf{S.Sem} & \textbf{S.Eq} & \textbf{A.Corr} & \textbf{A.Ver} & \textbf{SR} & \textbf{SR$\ge$0.5} & \textbf{FR} & \textbf{P} & \textbf{R} & \textbf{F1} & \textbf{P} & \textbf{R} & \textbf{F1} \\
        \hline
        \textbf{No Context} & -- & 0.86 & 0.56 & 0.55 & 0.96 & 0.98 & 0.66 & 0.53 & 0.76 & 44.0\% & 62.0\% & 38.0\% & 0.49 & 0.78 & 0.56 & 0.58 & 0.56 & 0.55 \\
        \textbf{Schema Only} & -- & 0.99 & 0.67 & 0.65 & 0.99 & 0.97 & 0.76 & 0.64 & 0.90 & 46.0\% & 82.0\% & 18.0\% & 0.74 & 0.86 & 0.78 & 0.57 & 0.62 & 0.53 \\
        \textbf{Schema + Graph} & -- & 0.99 & 0.70 & 0.72 & 0.99 & 0.99 & 0.79 & 0.70 & 0.94 & 56.0\% & 84.0\% & 16.0\% & 0.76 & 0.85 & 0.78 & 0.61 & 0.69 & 0.61 \\
        \textbf{Schema + Examples} & -- & 1.00 & 0.71 & 0.72 & 0.99 & 0.99 & 0.75 & 0.72 & 0.98 & 58.0\% & 86.0\% & 14.0\% & 0.78 & \textbf{0.88} & 0.80 & \textbf{0.66} & \textbf{0.75} & \textbf{0.66} \\
        \textbf{Full Context} & -- & 1.00 & \textbf{0.73} & \textbf{0.77} & 0.99 & 0.99 & 0.79 & \textbf{0.76} & 0.98 & \textbf{60.0\%} & \textbf{92.0\%} & \textbf{8.0\%} & \textbf{0.79} & 0.87 & \textbf{0.81} & 0.63 & 0.73 & 0.63 \\
        \hline
    \end{tabular}
    \caption{Comprehensive Context Ablation Matrix for \textbf{Claude 4.6 Sonnet}.}
    \label{tab:ablation_sonnet}
\end{table*}

\begin{table*}[ht!]
    \centering
    \scriptsize
    \setlength{\tabcolsep}{2.5pt}
    \begin{tabular}{l | cccc | c | cc | ccccc | ccc | ccc}
        \hline
        \textbf{Context} & \multicolumn{4}{c|}{\textbf{Retrieval \& Context Use}} & \textbf{Reasoning} & \multicolumn{2}{c|}{\textbf{SQL Quality}} & \multicolumn{5}{c|}{\textbf{Answer Quality}} & \multicolumn{3}{c|}{\textbf{Structural Tables}} & \multicolumn{3}{c}{\textbf{Structural Joins}} \\
        \textbf{Configuration} & \textbf{C.Rel} & \textbf{R.Use} & \textbf{C.Ev} & \textbf{E.Comp} & \textbf{R.Ut} & \textbf{S.Sem} & \textbf{S.Eq} & \textbf{A.Corr} & \textbf{A.Ver} & \textbf{SR} & \textbf{SR$\ge$0.5} & \textbf{FR} & \textbf{P} & \textbf{R} & \textbf{F1} & \textbf{P} & \textbf{R} & \textbf{F1} \\
        \hline
        \textbf{No Context} & -- & 0.59 & 0.58 & 0.58 & 0.99 & 0.99 & 0.63 & 0.58 & 0.74 & 44.0\% & 72.0\% & 28.0\% & 0.63 & 0.78 & 0.66 & 0.62 & 0.55 & 0.57 \\
        \textbf{Schema Only} & -- & 0.95 & 0.76 & 0.70 & 1.00 & 0.99 & 0.79 & 0.68 & 0.92 & 46.0\% & 90.0\% & 10.0\% & 0.76 & 0.86 & 0.78 & 0.61 & 0.66 & 0.59 \\
        \textbf{Schema + Graph} & -- & 0.96 & 0.68 & 0.67 & 1.00 & 0.99 & 0.75 & 0.73 & 0.94 & 50.0\% & 96.0\% & 4.0\% & 0.69 & 0.86 & 0.74 & 0.54 & 0.66 & 0.55 \\
        \textbf{Schema + Examples} & -- & 1.00 & 0.76 & 0.76 & 1.00 & 1.00 & 0.74 & 0.76 & 0.94 & 58.0\% & 94.0\% & 6.0\% & \textbf{0.81} & \textbf{0.90} & \textbf{0.83} & \textbf{0.69} & \textbf{0.76} & \textbf{0.69} \\
        \textbf{Full Context} & -- & 1.00 & \textbf{0.77} & 0.76 & 1.00 & 1.00 & \textbf{0.80} & \textbf{0.78} & 0.94 & \textbf{60.0\%} & 96.0\% & 4.0\% & 0.80 & 0.89 & 0.82 & 0.67 & 0.75 & 0.66 \\
        \hline
    \end{tabular}
    \caption{Comprehensive Context Ablation Matrix for \textbf{Claude 4.8 Opus}.}
    \label{tab:ablation_opus}
\end{table*}



\subsubsection{Accuracy vs. Efficiency}
\begin{figure}[ht!]
    \centering
    \includegraphics[width=\linewidth]{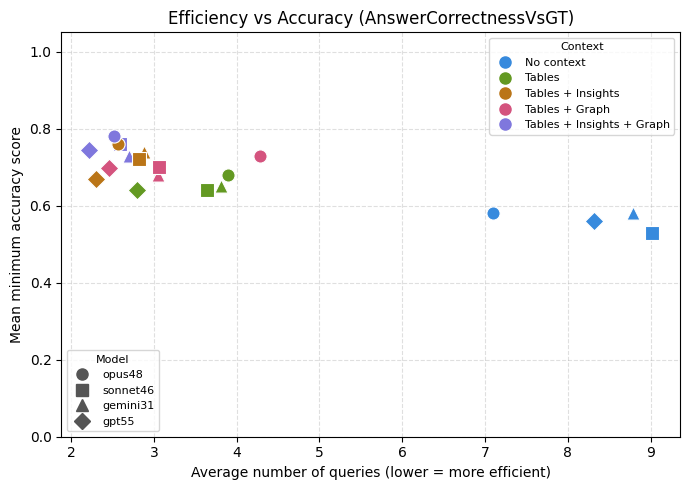} 
    \caption{Efficiency versus Accuracy (\texttt{AnswerCorrectnessVsGT}).}
    \label{fig:efficiency_vs_accuracy}
\end{figure}

Beyond answer quality, we additionally evaluate operational efficiency within the live agentic pipeline. Specifically, we measure how effectively different foundation models leverage the injected context to minimize iterative SQL exploration while preserving semantic correctness. Figure \ref{fig:efficiency_vs_accuracy} visualizes the resulting tradeoff between execution efficiency - measured as the average number of SQL queries executed per task - and final semantic accuracy (\texttt{AnswerCorrectnessVsGT}).

Across all evaluated architectures, richer contextual grounding consistently improves both efficiency and final accuracy. Moving from the \textbf{No Context} baseline to the strongest \textbf{Full Context} configuration reduces exploratory SQL queries by roughly 70\%, while improving \texttt{AnswerCorrectnessVsGT} by up to 34\% relatively (0.58 vs.\ 0.78 in Opus 4.8). In contrast, models without contextual grounding rely heavily on trial-and-error exploration and consistently achieve weaker semantic performance, highlighting the importance of explicit relational context for reliable cross-vendor identity reasoning.
As progressively richer metadata layers are injected, the models consistently migrate toward the top-left quadrant, demonstrating that structural context simultaneously improves both reasoning fidelity and execution efficiency. Importantly, the figure also reveals distinct architectural differences in how each foundation model takes advantage of contextual guidance. Claude 4.8 Opus establishes the strongest overall Pareto frontier, achieving the highest semantic accuracy while maintaining low query counts under the \textbf{Full Context} and \textbf{Tables + Insights} configurations. This behavior suggests that larger reasoning-oriented architectures are particularly effective in internalizing graph topology and relational abstractions into stable execution plans.

Claude 4.6 Sonnet demonstrates a different but equally important pattern: while its zero-context baseline lags behind Opus 4.8, the introduction of graph and example-driven context dramatically compresses the efficiency gap. Under richer contextual settings, Sonnet converges toward Opus-level operational efficiency while substantially improving semantic reliability, indicating that high-quality structural guidance can partially compensate for reduced intrinsic reasoning capacity.
Gemini 3.1 exhibits strong sensitivity to contextual enrichment, benefiting significantly from graph-aware reasoning augmentation. However, unlike the Claude family, Gemini appears to reach a mild context-saturation threshold where additional metadata layers yield diminishing semantic returns despite continued improvements in relational execution fidelity. This suggests that some architectures may struggle to optimally synthesize heterogeneous context modalities simultaneously.
Interestingly, GPT 5.5 consistently achieves some of the lowest average query counts across the benchmark, indicating highly efficient execution convergence. However, this efficiency is accompanied by lower semantic accuracy than the top-performing Claude configurations, suggesting a tradeoff between execution efficiency and deeper relational reasoning.

Collectively, these findings demonstrate that the value of the \textit{Sola} context engine extends beyond simple prompt enrichment. By explicitly injecting cross-vendor relational topology, schema structure, and retrieval exemplars, \textit{Sola} fundamentally reshapes the reasoning dynamics of frontier LLM agents—transforming inefficient exploratory behavior into targeted, graph-aware execution strategies capable of achieving both higher accuracy and lower computational overhead.

\subsubsection{Additional Insights}

In the Reasoning and Retrieval \& Context Use dimensions, the data reveals a fascinating behavioral trait of frontier models. Even under the zero-context baseline, the \texttt{ReasoningUtility} scores remain exceptionally high (ranging from 0.94 to 0.99 across all engines). This indicates that the models possess robust innate logical faculties; even when completely starved of database schema, they still attempt to formulate coherent, step-by-step retrieval plans. However, without Sola's explicit structural guidance, these logical plans are executed against hallucinated tables and incorrect entity assumptions. Once Sola injects the foundational metadata, \texttt{RetrievalUseQuality} immediately spikes to near-perfect scores (frequently hitting 0.99 or 1.00). This confirms that the models successfully abandon their baseline hallucinations, actively anchoring their high reasoning capabilities to Sola's provided schemas and few-shot examples to build a deterministic execution plan.

The SQL Quality and Structural metric tracks highlight the critical difference between generating plausible SQL and generating accurate domain-specific SQL. Even with basic schemas provided, generating true alignment with the ground truth - measured by \texttt{SQLEquivalenceVsGT} and the \texttt{JoinsF1} scores - only peaks when Sola introduces its advanced structural graphs and examples. Sola bridges the gap between basic table awareness and complex, multi-vendor entity resolution, proving that high-fidelity cybersecurity querying requires explicit relational mapping, not just basic table definitions.

A particularly important pattern emerges when comparing the relationship between \texttt{AnswerVerdictCorrectnessVsGT} and the stricter \texttt{AnswerCorrectnessVsGT} metric. Across nearly all evaluated architectures and context configurations, verdict correctness consistently establishes a substantially higher upper bound than full semantic correctness. This indicates that frontier LLMs are frequently capable of identifying the correct high-level security posture. Namely, correctly determining whether a configuration is compliant, risky, or misconfigured, while simultaneously failing to recover the complete supporting evidence required to fully justify that conclusion.

For example, under the \textbf{Full Context} configuration, Claude 4.8 Opus achieves an exceptionally high \texttt{AnswerVerdictCorrectnessVsGT} score of 0.94 while its stricter \texttt{AnswerCorrectnessVsGT} score remains lower at 0.78. Similar gaps appear consistently across GPT-5.5, Gemini 3.1 Pro, and Claude 4.6 Sonnet. This discrepancy reveals that the primary failure mode of modern frontier models is not necessarily high-level security reasoning itself, but rather evidence reconstruction fidelity: correctly enumerating all affected identities, recovering the precise supporting entities, reproducing exact join paths, and grounding conclusions in complete relational evidence.

This interpretation is further reinforced by the behavior of the \texttt{ContextualEvidenceRelevancy} and \texttt{EvidenceCompletenessVsGT} metrics. As richer graph-aware context is introduced, the gap between verdict-level correctness and full semantic correctness narrows substantially, demonstrating that the \textit{Sola} agent primarily improves the evidentiary grounding layer of the reasoning pipeline rather than merely teaching the models abstract security concepts. In practice, this distinction is operationally critical: an enterprise security agent that correctly identifies the existence of a privilege escalation path but fails to enumerate all impacted identities or misconfigured resources remains insufficient for production-grade remediation workflows.

Ultimately, the derived \textit{Answer Quality} aggregations serve as the final proof of work. The massive reduction in the \texttt{FailureRate} metric across all models under the Full Context setting demonstrates that Sola effectively stabilizes the agentic loop. Rather than leaving unassisted LLMs to navigate complex schemas blindly, which often results in a fragile, experimental generation process prone to mapping incorrect tables or arriving at wrong logical conclusions, Sola provides the structural rails needed to anchor the execution. This optimization successfully transforms a high-variance retrieval pipeline into a robust, production-ready data exploration engine.

\section{Conclusion}
\label{sec:conclusion}
In this work, we introduced the Cross-Vendor Sola ISPM Benchmark alongside a context orchestration framework for evaluating agentic AI systems on federated identity security reasoning tasks. Our empirical evaluation across Claude (Opus 4.8, Sonnet 4.6), ChatGPT 5.5, and Gemini 3.1 Pro demonstrates that \textbf{domain specific agentic enrichment and  orchestration layer improves performance significantly}, Unassisted LLMs are insufficient for reliable enterprise data exploration. By systematically ablating the context layers, we established that injecting rich metadata - relational schemas, semantic graphs, and few-shot exemplars - fundamentally stabilizes text-to-SQL execution. The \textit{Sola} agent acts as both an accuracy multiplier and an operational accelerator; it maximizes computational efficiency by empowering models to map complex joins on their initial attempts, drastically reducing the number of iterative queries and overall execution latency.

While our analysis revealed architectural nuances - Claude models maximized performance using the complete context payload, whereas Gemini exhibited a minor context-saturation threshold - the integration of structural graph topologies proved universally essential.

Ultimately, we demonstrate that an optimal balance of text-to-SQL accuracy and runtime efficiency is achieved through precise, domain-aware context orchestration. Future work will extend this work to include more harder multi hop questions (specifically 4 hops and more) and broader ISPM evaluation suite such as, behavioral analytics, identity risk scoring, mitigation planning and governance alignment.
\bibliographystyle{unsrt}  
\bibliography{references}

\appendix
\section{Benchmark Question Set}

To thoroughly assess the capability of agentic AI systems in correlating identity security posture across heterogeneous environments, we evaluated a comprehensive set of cross-vendor visibility questions. These queries simulate advanced threat hunting, offboarding verification, and compliance tasks that require synthesizing data from multiple identity providers, cloud infrastructure services, and SaaS applications. The following groupings detail the experimental questions, categorized by their target vendor combinations to demonstrate the required interoperability breadth.  \paragraph{AWS and Okta}
\begin{quote}
\vspace{-1em}
\textit{\begin{enumerate}
\item Which AWS SSO users with admin-level access have an inactive or missing Okta account? \item Can anyone without MFA can access our production systems S3 Buckets or AWS Databases? \item Which AWS SSO users haven't logged into Okta in 90 days but are still active in Okta? \item Are there any users that are deactivated in Okta but still active in AWS? \item List AWS IAM Users who have an active Access Key (Secret) older than 90 days but have never logged into Okta. \item Which active AWS SSO users in Okta have no phishing-resistant MFA factor enrolled? 
\end{enumerate}}
\end{quote}  \paragraph{HiBob and Okta}
\begin{quote}
\vspace{-1em}
\textit{\begin{enumerate}
\item Are there any Okta accounts that don't match a current employee or contractor in HiBoB? \item Which employees have active admin role assignments in Okta and what is their department in HiBob? 
\end{enumerate}}
\end{quote}  \paragraph{Google Workspace and Okta}
\begin{quote}
\vspace{-1em}
\textit{\begin{enumerate}
\item Who are the Google Workspace admins that aren't listed as admins in Okta? \item Which users haven't used GWS for 90 days but still listed as active in Okta? \item Are there any disabled Okta users who still have active Google Workspace accounts? \item Which active Google Workspace users have role assignments, and do they have a matching Okta identity (including Okta status)? \item List all users who have an active account in Google Workspace but are synced through Okta. \item Which Okta users have the most publicly accessible files in Google Workspace? \item Are there any external email addresses with access to any shared GWS files that are not managed through Okta? 
\end{enumerate}}
\end{quote}  \paragraph{GitHub and Okta}
\begin{quote}
\vspace{-1em}
\textit{\begin{enumerate}
\item Are there any GitHub org members not provisioned through Okta? \item Which GitHub users are Organization Owners but do not have a corresponding active account in Okta? \item Are there deprovisioned users in Okta with admin access in GitHub? \item Which active Okta users who are GitHub organization members have two-factor authentication disabled on GitHub, broken down by their organization role (MEMBER vs ADMIN)? 
\end{enumerate}}
\end{quote}  \paragraph{HiBob and AWS}
\begin{quote}
\vspace{-1em}
\textit{\begin{enumerate}
\item Do any offboarded employees still have active AWS accounts? 
\end{enumerate}}
\end{quote}  \paragraph{HiBob and GitHub}
\begin{quote}
\vspace{-1em}
\textit{\begin{enumerate}
\item Show me users who left the company but still have GitHub access. \item Are any marketing or finance users granted write access to our code base? \item Are there any GitHub org members with no matching HiBob employee record? \item Which employees marked as terminated in HiBob have authored commits in GitHub after their recorded termination date? 
\end{enumerate}}
\end{quote}  \paragraph{HiBob and Google Workspace}
\begin{quote}
\vspace{-1em}
\textit{\begin{enumerate}
\item Do any non-IT or non-R\&D employees have super admin access to Google Workspace? \item Which former employees still own Google Workspace documents (private or shared drives) after their termination? \item Do any terminated employees own publicly accessible files in their Google Drive personal storage? \item Are there any terminated employees who have both an active Google Workspace account and direct user-level IAM role bindings in our GCP projects? \item Which terminated users still have write permissions to any shared GWS documents not owned by them? \item Which terminated users still have read permissions to any shared GWS documents not owned by them? 
\end{enumerate}}
\end{quote}  \paragraph{Azure AD and Okta}
\begin{quote}
\vspace{-1em}
\textit{\begin{enumerate}
\item Who are the Azure AD admins that aren't listed as admins in Okta? 
\end{enumerate}}
\end{quote}  \paragraph{MongoDB Atlas and Okta}
\begin{quote}
\vspace{-1em}
\textit{\begin{enumerate}
\item Who has admin access in Mongo that wasn't provisioned through Okta? 
\end{enumerate}}
\end{quote}  \paragraph{GitHub and Azure AD}
\begin{quote}
\vspace{-1em}
\textit{\begin{enumerate}
\item Are there any active GitHub accounts for users who have been disabled in Azure AD? \item Which GitHub organization members have no corresponding identity in Azure AD? 
\end{enumerate}}
\end{quote}  \paragraph{Google Workspace and GCP}
\begin{quote}
\vspace{-1em}
\textit{\begin{enumerate}
\item Which Google Workspace users with GCP IAM permissions can bypass MFA? \item Which Google Workspace administrators also hold a primitive Owner or Editor role in GCP? 
\end{enumerate}}
\end{quote}  \paragraph{GitHub and GCP}
\begin{quote}
\vspace{-1em}
\textit{\begin{enumerate}
\item Which GitHub organization members with two-factor authentication disabled also hold direct user-level IAM role bindings in GCP, and what privileged roles do they have? \item Which GitHub organization members hold GCP primitive roles (Owner or Editor), and what is their GitHub role? 
\end{enumerate}}
\end{quote}  \paragraph{Azure AD and Google Workspace}
\begin{quote}
\vspace{-1em}
\textit{\begin{enumerate}
\item Which users exist in both EntraID and GWS but don't have 2SV enforced in Google Workspace? 
\end{enumerate}}
\end{quote}  \paragraph{HiBob, Okta, and AWS}
\begin{quote}
\vspace{-1em}
\textit{\begin{enumerate}
\item Are there any AWS IAM credentials whose owning IAM user cannot be traced back to an active HiBob employee through Okta? \item According to Hibob, do any employees classified as contractors, external, or freelancers have AWS SSO account? \item Which users have AWS SSO access but don't have a corresponding employee record in our Hibob system? \item Which terminated employees still have AWS SSO access? 
\end{enumerate}}
\end{quote}  \paragraph{HiBob, Okta, and Google Workspace}
\begin{quote}
\vspace{-1em}
\textit{\begin{enumerate}
\item Are there any Google Workspace accounts still active for employees who have been terminated? 
\end{enumerate}}
\end{quote}  \paragraph{HiBob, Okta, and MongoDB Atlas}
\begin{quote}
\vspace{-1em}
\textit{\begin{enumerate}
\item Are there any MongoDB Atlas users who are terminated in HiBob or deactivated in Okta? 
\end{enumerate}}
\end{quote}  \paragraph{HiBob, Okta, and GitHub}
\begin{quote}
\vspace{-1em}
\textit{\begin{enumerate}
\item Which terminated external workers (non-Sola employees) in HiBob still have an active Okta account or GitHub organization membership? 
\end{enumerate}}
\end{quote}  \paragraph{GitHub, AWS, and MongoDB Atlas}
\begin{quote}
\vspace{-1em}
\textit{\begin{enumerate}
\item Which GitHub organization members without MFA enabled also have AWS admin-level permissions or MongoDB Atlas admin roles (ORG\_OWNER, ORG\_BILLING\_ADMIN, GROUP\_OWNER)?
\end{enumerate}}
\end{quote}  \paragraph{Azure AD, Okta, and GitHub}
\begin{quote}
\vspace{-1em}
\textit{\begin{enumerate}
\item Which Azure AD users registered only with weak MFA methods (SMS or phone call) are also GitHub organization members? 
\end{enumerate}}
\end{quote}  \paragraph{Okta, AWS, GitHub, and Google Workspace}
\begin{quote}
\vspace{-1em}
\textit{\begin{enumerate}
\item Which active Okta users are provisioned in AWS Identity Store, hold ADMIN permission on at least one GitHub repository, and have a Google Workspace administrator role? 
\end{enumerate}}
\end{quote}  \paragraph{HiBob, Okta, Google Workspace, Azure AD, and AWS SSO}
\begin{quote}
\vspace{-1em}
\textit{\begin{enumerate}
\item Which terminated employees have not been fully offboarded from all connected systems (Okta, Google Workspace, Azure AD, or AWS SSO)? 
\end{enumerate}}
\end{quote}




\section{Evaluation Criteria and Rubrics}
\label{app:evaluation-criteria}

This appendix provides the complete criteria prompts and rubrics used in the 
LLM-as-Judge evaluation of the benchmark. For each metric, 
we include: (1) a short description of what is evaluated, (2) the exact criteria 
prompt provided to the judging model, and (3) the scoring rubric. All metrics 
use a three-level scale: 0 (Does Not Meet Criterion), 0.5 (Partially Meets 
Criterion), and 1 (Fully Meets Criterion).

\subsection{Retrieval Use Quality}

\paragraph{What is evaluated.}
This metric evaluates how effectively the agent utilizes retrieved examples, schema information, and prior patterns when constructing SQL queries for the current ISPM task. Rather than measuring retrieval relevance itself, the metric focuses on whether the agent successfully adapts the retrieved context to the current question and schema.

\paragraph{Criteria.}
\begin{quote}\ttfamily\raggedright
You are a senior expert in the domain Identity Security Posture Management (ISPM),
specializing in Questions about identities (human and non-human), their privileges,
access paths, authentication posture (MFA, phishing-resistant methods), lifecycle hygiene
(onboarding/offboarding, dormant accounts), policy/configuration posture, and resulting
identity-centric risk across cloud and SaaS environments.

You are evaluating HOW WELL the agent UTILIZES retrieved patterns (examples) and schemas
when constructing SQL for the current QUESTION.

You will see:

- QUESTION (what needs to be answered).\\
- ACTUAL\_OUTPUT: the concatenated SQL\_QUERIES generated by the agent.\\
- CONTEXT, which may include: \\ 
• SCHEMAS: authoritative structures for tables/columns. \\ 
• RETRIEVED\_EXAMPLES: historical questions/answers/SQL patterns. \\ 
• STEP\_JOURNAL: reasoning and tool calls.

Evaluate whether the agent correctly identifies and adapts the most relevant examples and schemas to the current task, while avoiding irrelevant template reuse or schema misuse.
\end{quote}

\paragraph{Rubric.}
\begin{quote}\ttfamily\raggedright
0.0 — Poor utilization: SQL largely ignores clearly useful examples/schemas or misuses them through irrelevant templates or invalid schema references.

0.5 — Partial utilization: SQL demonstrates some adaptation of retrieved examples and schemas, but contains gaps, missed opportunities, or inconsistent pattern usage.

1.0 — Good utilization: SQL appropriately adapts the most relevant retrieved patterns and schemas to the current question while avoiding irrelevant or invalid template reuse.
\end{quote}

\subsection{Contextual Relevancy}

\paragraph{What is evaluated.}
This metric evaluates whether the retrieval candidates selected by the agent are relevant to the current ISPM question. The focus is on the quality of the retrieved context itself rather than the final reasoning or answer quality.

\paragraph{Criteria.}
\begin{quote}\ttfamily\raggedright
You are a senior expert in the domain Identity Security Posture Management (ISPM).

You are evaluating how relevant the agent's reranked retrieval candidates are to the QUESTION.

CONTEXT contains the final reranked retrieval candidates consumed by the agent. Each candidate may contain descriptions, schemas, SQL fragments, or related entities.

Judge whether the retrieved candidates are genuinely useful for answering the current QUESTION.

Reward candidates that directly address the entities, concepts, tables, or security relationships relevant to the question. Penalize off-topic, redundant, or unrelated retrieval results.
\end{quote}

\paragraph{Rubric.}
\begin{quote}\ttfamily\raggedright
0.0 — Mostly irrelevant: retrieved candidates do not meaningfully address the entities or concepts required for the question.

0.5 — Mixed relevance: some candidates are relevant, but many are redundant, off-topic, or unrelated.

1.0 — Highly relevant: retrieved candidates clearly align with the entities, tables, and security concepts required to answer the question.
\end{quote}

\subsection{Contextual Evidence Relevancy}

\paragraph{What is evaluated.}
This metric evaluates whether the entities and evidence retrieved by the agent match the evidence retrieved by the ground-truth SQL execution. It measures evidence overlap rather than answer correctness.

\paragraph{Criteria.}
\begin{quote}\ttfamily\raggedright
You are evaluating whether the entities and evidence fetched by the agent match the entities and evidence fetched by the ground-truth execution for the same QUESTION.

CONTEXT contains SQL results generated by both the agent and the ground truth.

Compare the returned entities, identifiers, statuses, counts, and security-relevant evidence between the two outputs.

Paraphrasing, row ordering, and formatting differences do not matter.
\end{quote}

\paragraph{Rubric.}
\begin{quote}\ttfamily\raggedright
0.0 — Mismatch: agent results miss key ground-truth entities or contradict the ground truth.

0.5 — Partial match: some ground-truth entities are recovered, but important entities, statuses, or counts are missing.

1.0 — Strong match: agent results successfully recover the key entities and supporting evidence present in the ground truth.
\end{quote}

\subsection{Evidence Completeness vs Ground Truth}

\paragraph{What is evaluated.}
This metric evaluates the completeness of the supporting evidence provided by the model answer relative to the supporting evidence contained in the ground-truth answer.

\paragraph{Criteria.}
\begin{quote}\ttfamily\raggedright
You are comparing the supporting context in a model answer to the supporting context in a ground-truth answer.

ACTUAL\_OUTPUT is the model answer. EXPECTED\_OUTPUT is the ground-truth answer.

Evaluate factual overlap of supporting evidence such as named users, accounts, repositories, permissions, systems, statuses, counts, and explanatory qualifiers.

Paraphrases are acceptable. Ordering and formatting differences do not matter.
\end{quote}

\paragraph{Rubric.}
\begin{quote}\ttfamily\raggedright
0.0 — Poor evidence match: little overlap with ground-truth supporting evidence or major contradictory evidence.

0.5 — Partial evidence match: some important supporting evidence is present, but material entities or qualifiers are missing.

1.0 — Strong evidence match: the model captures the material supporting evidence contained in the ground truth without major contradictions.
\end{quote}

\subsection{Reasoning Utility}

\paragraph{What is evaluated.}
This metric evaluates whether the agent's reasoning trajectory makes meaningful progress toward solving the ISPM question. The focus is not on correctness alone, but on whether the reasoning path investigates the appropriate entities, tables, and sub-questions needed for the task.

\paragraph{Criteria.}
\begin{quote}\ttfamily\raggedright
You are evaluating whether the agent's reasoning steps make plausible progress toward the correct answer for the QUESTION.

CONTEXT includes: \\
- SCHEMAS \\ 
- STEP\_JOURNAL

A reasoning step has utility when it investigates entities, tables, joins, or sub-questions that plausibly contribute to solving the task.

Penalize reasoning paths that pursue irrelevant entities, ignore the primary data source required by the question, or investigate unrelated security concepts.
\end{quote}

\paragraph{Rubric.}
\begin{quote}\ttfamily\raggedright
0.0 — No utility: reasoning investigates irrelevant entities or misses the primary data sources required to answer the question.

0.5 — Partial utility: reasoning explores some relevant concepts but also contains substantial detours or omissions.

1.0 — Strong utility: reasoning consistently targets the correct entities, tables, joins, and concepts required for solving the task.
\end{quote}

\subsection{SQL Semantic Appropriateness}

\paragraph{What is evaluated.}
This metric evaluates whether the generated SQL queries represent a semantically appropriate strategy for solving the ISPM question, independent of exact equivalence to the ground-truth SQL.

\paragraph{Criteria.}
\begin{quote}\ttfamily\raggedright
You are evaluating the semantic appropriateness of SQL queries generated for an ISPM question.

ACTUAL\_OUTPUT contains the generated SQL queries.

Determine whether the SQL structure, joins, filters, and data access patterns plausibly solve the intended task while remaining consistent with the available schema.
\end{quote}

\paragraph{Rubric.}
\begin{quote}\ttfamily\raggedright
0.0 — Inappropriate: SQL is structurally incorrect, inconsistent with the schema, or unable to solve the task.

0.5 — Partially appropriate: SQL captures some correct logic but contains important structural or semantic issues.

1.0 — Appropriate: SQL represents a coherent and semantically correct strategy aligned with both the schema and question intent.
\end{quote}

\subsection{SQL Equivalence vs Ground Truth}

\paragraph{What is evaluated.}
This metric evaluates whether the model-generated SQL plan is logically equivalent to the ground-truth SQL plan used to derive the benchmark answer.

\paragraph{Criteria.}
\begin{quote}\ttfamily\raggedright
You are comparing model-generated SQL against ground-truth SQL.

ACTUAL\_OUTPUT is the model-generated SQL plan. EXPECTED\_OUTPUT is the ground-truth SQL plan.

Judge whether both SQL plans implement equivalent logic, including similar joins, filters, aggregations, and multi-step reasoning structure.
\end{quote}

\paragraph{Rubric.}
\begin{quote}\ttfamily\raggedright
0.0 — Not equivalent: SQL logic substantially differs from the ground-truth approach or cannot solve the task correctly.

0.5 — Partially equivalent: SQL shares some logical structure with the ground truth but contains missing or incorrect logic.

1.0 — Equivalent: SQL plans are logically equivalent approaches for solving the task.
\end{quote}

\subsection{Answer Verdict Correctness vs Ground Truth}

\paragraph{What is evaluated.}
This metric evaluates whether the model reaches the same primary yes/no security verdict as the ground-truth answer, independent of supporting details.

\paragraph{Criteria.}
\begin{quote}\ttfamily\raggedright
You are comparing a model answer to a ground-truth answer.

Evaluate ONLY whether the model reaches the same primary yes/no verdict as the ground truth.

Ignore supporting entities, counts, ordering, or additional context unless they explicitly contradict the verdict.
\end{quote}

\paragraph{Rubric.}
\begin{quote}\ttfamily\raggedright
0.0 — Incorrect verdict: the model reaches the opposite or contradictory yes/no conclusion.

1.0 — Correct verdict: the model reaches the same primary yes/no conclusion as the ground truth.
\end{quote}

\subsection{Answer Correctness vs Ground Truth} 

\paragraph{What is evaluated.} This metric determines whether the agent's final answer reaches the same security-relevant conclusions as the ground-truth answer, regardless of phrasing. In ISPM this includes privilege classification, lifecycle hygiene interpretation, exposure determination, authentication posture, and identity risk assessments. 

\paragraph{Criteria.} 
\begin{quote}\ttfamily\raggedright 
You are a senior expert in the domain Identity Security Posture Management (ISPM), specializing in Questions about identities (human and non-human), their privileges, access paths, authentication posture (MFA, phishing-resistant methods), lifecycle hygiene (onboarding/offboarding, dormant accounts), policy/configuration posture, and resulting identity-centric risk across cloud and SaaS environments. You are comparing a model answer to a ground-truth answer. ACTUAL\_OUTPUT is the model's answer. EXPECTED\_OUTPUT is the ground-truth answer. Judge whether they express the same security-relevant conclusion (entities,classifications, posture, counts, etc.). \end{quote}

\paragraph{Rubric.}
\begin{quote}\ttfamily\raggedright 0.0 — Incorrect: model answer reaches a different or wrong conclusion. 0.5 — Partially correct: same direction as GT but different magnitude (e.g., GT says exactly 18, model says 'at least 5' or 'about 10'), or some overlap with important differences or omissions. 1.0 — Correct: model answer is semantically equivalent to the ground truth. Hedged counts ('at least N', 'around N') are acceptable as Correct only if (a) N is consistent with the GT count ($\ge$ GT for 'at least', within $\pm$ 10\% for 'around'), and (b) the named entities are a correct subset of the GT entities. A hedged count is NOT correct if the listed items are wrong or if the magnitude differs substantially. \end{quote}

\end{document}